\documentclass[aps,prd,onecolumn,showpacs]{revtex4}
\usepackage{epsfig}
\def\planck{{\sl Planck }}
\def\wmap{{\sl WMAP }}
\newcommand{\bi}[1]{\mbox{\boldmath $#1$}}
\def\l{{\ell}}
\def\etal{{et al.}}

\newcommand{\lm}{{\l m}}
\def\alm{a_{\l m}}
\def\Ylm{Y_{\l m}}
\def\cl{C_{\l}}

\def\summ{\sum_{m=-\l}^{\l}}
\def\suml{\sum_{\l=0}^{\infty}}

\def\healpix{H{\sc ealpix }}

\def\wmap{\hbox{\sl WMAP~}}

\def\etal{et al.}
\def\alm{a_{\l m}}
\def\Ylm{Y_{\l m}}
\def\Cl{C_{\l}}

\def\summ{\sum_{m=-\l}^{\l}}
\def\suml{\sum_{\l=0}^{\infty}}
\def\lm{{\l m}}
\def\a{{\bf a}}

\def\C{{\bf C}}
\def\S{{\bf S}}

\def\R{{\bf R}}

\def\S{{\bf S}}

\def\lmax{\l_{\max}}

\def\planck{{\it Planck }}

\begin{document}
\title{Statistics of phase correlations as a test for non-Gaussianity of the CMB maps}
\author{Pavel Naselsky, Lung-Yih Chiang, Poul Olesen and Igor Novikov}
\affiliation{Niels Bohr Institute, Blegdamsvej 17, DK-2100,  Copenhagen, Denmark}
\date{\today}
\begin{abstract}
Gaussianity is the very base for derivation of the cosmological parameters from the CMB angular power spectrum. Non-Gaussian signal, whether originated from experimental error or primordial source, could mimic extra power in the power spectrum, thereby leading to a wrong set of parameters. In this paper we present a new Gaussianity analysis of the derived CMB singals. It is based on the trigonometric moments of phases, which can be concluded with a ``mean angle'',  allowing us to see the global trend of non-Gaussianity of the signals. These moments are also  closely related to the Pearson's random walks. We apply these analyses on the derived CMB maps and their different morphologies manifest themselves through these functions. We also comment on rotational invariance of the trigonometric moments of phases as a non-Gaussianity test.

\end{abstract}
\pacs{PACS numbers: 98.80.Hw, 52.34.Mw}

\maketitle
\section{Introduction}
One of the main goals of the ESA \planck mission is to produce the maps of CMB temperature anisotropies and polarization and the power spectra, from which the accurate values of the cosmological parameters can be derived such as baryonic and dark matter density $\Omega_m$, the dark energy density $\Omega_{d}$, the optical depth of reionization $\tau_r$ and so on. A crucial requirement that enables us to derive the cosmological parameters from the temperature and polarization power spectra $C_T(\l)$ and $C_p(\l)$ is that the statistical properties of the primordial CMB signal should be Gaussian. Should the primordial CMB
signal possess non-Gaussian origin such as quadratic non-linearity in the gravitational
potential (Komatsu et al. 2003, see also Wandelt et al. 2004, Mattarese
et al. 2004), the connection between $C_T(\l)$, $C_p(\l)$ and the cosmological parameters
needs an additional investigation which seems to be non-trivial. One can illustrate the importance of non-Gaussianity of the CMB signal by assuming that at some range of multipoles, say, $\l \sim 200$ the $\alm$ coefficients of the spherical harmonics expansion
of the anisotropy $\Delta T$ are highly correlated.
Without comprehensive testing of non-Gaussianity of the map, these correlations can easily mimic the first acoustic peak in the $C_T(\l)$, leading to a wrong conclusion on the properties of the CMB and the cosmological parameters.
Preparation and implementation of sensitive non-Gaussianity tests on the anisotropy and polarization maps is therefore pivotal for the \planck mission.

After the release of 1-year results of the Wilkinson Microwave
Anisotropy Probe (\wmap) \cite{wmap,wmapresults,wmapfg,wmapsys,wmapcl}
in the papers by Chiang et al. (2003), Park (2004), Eriksen et al. (2004), Hansen et al. (2004), Larson and Wandelt (2004), Land and Magueijo (2004), Roukema et al. (2004), Schwarz et al. (2004), Chen et al. (2004), various kinds of non-Gaussian features are detected in
the \wmap derived maps. These structures of the \wmap signal in combination with 
phase analysis of the whole sky maps, derived from the \wmap data sets (Chiang et al. 2003, Naselsky, Doroshkevich and Verkhodanov 2004)
and wavelet analysis (Vielva et al. 2004, Cruz et al. 2005) clearly show that \wmap signal contains non-Gaussian features, which could be primordial origin (Eriksen et al. 2004) or related with
foreground residues (Naselsky et al. 2004, Chiang and Naselsky 2004, Dineen and Coles 2004). 
The exercise of finding non-Gaussian peculiarities in the \wmap CMB maps is also crucial for the upcoming \planck mission, paving the way especially for reconstruction of the signal in the Galactic plane area, in order to obtain the whole CMB sky for both temperature anisotropies and polarization. 
The method of testing non-Gaussianity presented in this paper operate in the multipole coefficients $\alm$ and related to phase analysis, developed by Chiang and Coles (2000), Chiang, Naselsky and Coles (2004), Coles et al. (2004), Naselsky et al. (2004a,b), Dineen, Graca and Coles (2005). It is applicable for both temperature anisotropies and E component of polarization multipole coefficients. 

The purpose of this paper is to show that any correlations of phases for CMB signal determine the morphology in the space of phases. We generalize the method of phase correlations by Naselsky, Doroshkevich and Verhodanov (2003) for testing of the phase coupling in the
CMB maps. Particularly, we introduce the mean angle $\Theta_\l$ for each multipole $\l$,
averaged over all $m$-modes in order to check whether the distribution of  $\Theta_\l$ is uniform (as it should be for Gaussian signals), or, if not, what possible preferable directions for each miltipole could be. Note that
this statistic of phases has a natural explanation in terms of Pearson's random
walk problem (Pearson 1905) for $\sum_m \alm$ and can be easily
generalized for more complex correlations of phases. The importance of such approach is recently pointed out by Stennard and Coles (2004).

For illustration of sensitivity of the method we use the foreground cleaned map (FCM) and  Wiener filtered map (WFM) derived by Tegmark, de Oliveira-Costa and Hamilton (2003) (hereafter TOH) foregrounds cleaned map (FCM), the ILC map by \wmap science team ( \footnote{$http://lambda.gsfc.nasa.gov/product/map/m_products.cfm$})
and the ILC map reproduced by Erikson et al. (2004) (hereafter EILC map). All these maps contain some features of the foreground residues, non-uniformity of the noise, Galactic plane substraction \ldots etc. To avoid confusion, below we call these maps the CMB maps, which represent different morphology of the
primordial (Gaussian) CMB signals with some of the non-Gaussian features. The main task is to
show how any detected non-Gaussian features relate to known properties of the
non-Gaussian components of the signal. All these maps include Galactic plane contamination, 
which is excluded by different kind of masks in order for estimation of the power spectrum of CMB anisotropies.

\section{Phase statistics and their application for the CMB anisotropy problem}
In this section we formulate the problem: how do the anisotropies of the CMB signal manifest themselves in the phase correlations? To answer this question, we recap some of the basics of Gaussian random field. The CMB temperature fluctuations on a sphere can be expressed as a sum over spherical harmonics:
\begin{equation}
\Delta T(\theta,\varphi)=\suml \summ |\alm|e^{i\phi_{lm}} \Ylm (\theta,\varphi),
\label{eq1}
\end{equation}
where $|\alm|$ and $\phi_{lm}$ are the moduli and phases of the coefficients of the expansion.

In practice, we use the \healpix package \cite{healpix} to decompose
each of the following derived CMB maps (ILC, FCM, WFM, EILC) for the coefficients of spherical harmonics $\alm$, extracting the phases for analysis.

A homogeneous and isotropic CMB Gaussian random field (GRF), as a result
of the simplest inflation paradigm, possess Legendre polynomial expansion modes whose real
and imaginary parts are Gaussian and mutually independent \cite{bbks,be}. The statistical
properties of such a field are then completely specified by its angular power spectrum
$\Cl^{cmb}$,
\begin{equation}
\langle  a^{cmb}_{\l^{ } m^{ }} (a^{cmb})^{*}_{\l^{'} m^{'}}
\rangle = \Cl^{cmb} \; \delta_{\l^{ } \l^{'}} \delta_{m^{} m^{'}}.
\label{eq2}
\end{equation}
In other words, the Central Limit Theorem guarantees that the field is Gaussian if their phases
\begin{equation}
\Psi^{cmb}_{\l m}=\tan^{-1}\frac{\Im (\alm^{cmb})}{\Re (\alm^{cmb})}
\label{eq3}
\end{equation}
are randomly and uniformly distributed at the range $[0,2\pi]$.
The method we propose is therefore based on the so-called ``random phase hypothesis'', our null hypothesis for a Gaussian random field, by testing any phase correlations between modes.  For this purpose we apply circular statistics \cite{fisher} on phases of the spherical harmonic coefficients (Naselsky et al. 2003; 2004). We examine the cross-correlation of phases between modes $(\l,m)$ and $(\l+ \Delta \l, m+\Delta m)$ for all $\l$, $m$ values. The cross-correlation of phases
vanishes by definition for pure Gaussian CMB signal, whlist for non-Gaussian components it should display non-trivial significance. Note that we do not examine any specific kind of non-Gaussian signals. In reality, all different sorts of non-Gaussian components can propagate into the data, such as systematic errors and foreground residues, in addition to primordial non-Gaussianity (if exists).

The basic idea is to introduce trigonometric moments of phases, which minimize the contribution of the non-correlated Gaussian tail and
maximize the non-Gaussian tail of the phases (Naselsky, Doroshkevich and Verkhodanov 2003;
Naselsky et al. 2004). To simplify the analysis let us describe the phase correlations in $\l-m$ plane separately, choosing orthogonal directions $\Delta_\l \neq 0$, $\Delta_m=0$ and $\Delta_\l= 0$, $\Delta_m \neq 0$ and throughout this paper we demonstrate phase coupling in these two directions. Following \cite{ndv03,ndv04} we define the following trigonometric moments
for the direction $\Delta_\l \neq 0, \Delta_m=0$:
\begin{eqnarray}
\C(\l,\Delta_\l)=\sum_{m=1}^{\l} \cos\left(\Psi_{\l+\Delta \l,m}-\Phi_{\l,m}\right); \nonumber\\
\S(\l,\Delta_\l)=\sum_{m=1}^{\l} \sin\left(\Psi_{\l+\Delta \l,m}-\Phi_{\l,m}\right); \nonumber\\
\label{def2}
\end{eqnarray}
and for the direction $\Delta_\l =0, \Delta_m \neq 0$
\begin{eqnarray}
\C(m,\Delta_m)=\sum_{\l=m}^{\l=\lmax} \cos\left(\Psi_{\l,m}-\Phi_{\l,m+\Delta m}\right); \nonumber\\
\S(m,\Delta_m)=\sum_{\l=m}^{\l=\lmax} \sin\left(\Psi_{\l,m}-\Phi_{\l,m-\Delta m}\right); \nonumber\\
\label{def2a}
\end{eqnarray}
where $\lmax$ is the maximum $\l$ putting into calculation, $\Psi_{\l m}$ and $\Phi_{\l m}$ can be the phases of the spherical harmonics of two maps for cross correlation, or those of the same map ($\Psi_{\l m}=\Phi_{\l m}$) for auto-correlation.
The concept behind such statistics is clear. If the phases of the signal are highly correlated ($\Psi_{\l,m} \rightarrow \Phi_{\l+\Delta \l,m})$, for the $\S_{NG}$
statistics we get $\S \rightarrow 0$, while $\C \rightarrow 1$. We therefore can further device the ``{\it mean angle}'' along $m$ and $\l$ direction as
\begin{eqnarray}
\Theta(\l,\Delta_\l)=\tan^{-1}\frac{\S(\l,\Delta_\l)}{\C(\l,\Delta_\l)};\nonumber\\
\Theta(m,\Delta_m)=\tan^{-1}\frac{\S(m,\Delta_m)}{\C(m,\Delta_m)},
 \label{asym}
\end{eqnarray}
respectively, both defined in $[0, 2\pi]$. Thus, for the pure Gaussian signals the mean angles should be uniformly random along either the $\l$ (after summation over $m$) or $m$ (over $\l$) direction. Non-Gaussian signals in the CMB data manifest themselves as asymmetry with respect to $\pi$ or non-random.
Moreover, for non-Gaussian signal we expect to find cross-correlation between $\Theta(\l,\Delta_\l)$ and $\Theta(\l^{'},\Delta_\l)$, and between
$\Theta(m,\Delta_m)$ and $\Theta(m^{'},\Delta_m)$ for different $\l,\l^{'}$ and $m,m^{'}$ modes.

\section{Trigonometric statistics as an example of Pearson's random walk}
One of the interesting properties of the trigonometric statistics is related to the idea suggested in Stannard and Coles (2004), in which it is pointed out that the statistics of the $\alm$ coefficients in the form of $\R_\l=\sum_{m>0} \alm$ is essentially the same as those of ``Rayleigh flight'', whose properties are well developed. However, for the trigonometric statistics proposed in the previous Section, we need to apply Pearson's random walk statistics, taking into account the investigation of the random walk with given (with fixed length of steps) by Pearson in his famous letter to the Journal Nature \cite{pearson}. Following Hughes (1995) and Stannard and Coles (2004), we introduce the random variable
\begin{equation}
G_\l(M)=\sum_{m=1}^{m=M}\frac{\alm}{|\alm|}=\sum_{m=1}^{m=M} e^{i \Psi_{\l m}},
\label{pear0}
\end{equation}
the real and imaginary parts of which are essentially following the properties of the Pearson's Walk in the complex plane after $M$ steps. Note that the walk can be in $\l$ direction for $G_m(M)$. For all the vectors $G(M)$ the starting point is zero and the length of each step is unity: $|a_\lm|=1$.  As one can see from Eq.(\ref{pear0}), vector $G_\l(M)$ has a projections on $x$ and $y$ axes $\sum\cos\Psi_{\lm}$ and $\sum\sin\Psi_{\lm}$, which are the trigonometric moments introduced above. The probability density function (PDF) for the position of the walks after $M=\l$ steps satisfies the equation (see Hughes (1995) for references)
\begin{equation}
P_{M+1}(\bi r )=\int p_{M+1}(\bi r-{\bi r ^{'}})P_{M}({\bi r^{'}})d^2{\bi r^{'}}
\label{pearb}
\end{equation}
where $p_{M}(\bi r)$ is the probability density function that the vector $G(M)$ lies in an infinitesimal area
 centered on
$\bi r$ on the $M$-th step. For isotropic two-dimensional random walk, in which each step has the same length
$a=1$ we get for $p_{M}(\bi r)$ (Kluyver 1906)
\begin{equation}
p_{M=1}(\bi r)=\frac{1}{2\pi|\bi r|}\delta(|\bi r| - 1)
\label{pearb1}
\end{equation}
and the corresponding PDF for $M$ steps is
\begin{equation}
 P_M(\bi r)=\frac{1}{2\pi}\int_0^{\infty} d\nu \; \nu J_0(\nu |\bi r|)\left[J_0(\nu)\right]^M,
\label{pear1}
\end{equation}
if $ M \ge 2$ and $P_1(\bi r)=p_{1}(\bi r)$ for $M=1$. If $M \gg 1$ the PDF has Gaussian asymptotic (see the references in Hughes (1995))
\begin{equation}
P_{M}({\bi r}) \simeq \frac{1}{\pi M}\exp\left(-\frac{|\bi r|^2}{M}\right).
\label{pear2}
\end{equation}
So, for $M\gg 1$ the corresponding variance of the random process is $\sigma^2= M/\sqrt 2$.
We would like to point out that this result can be obtained directly from the circular variables $\C$ and $\S$ by the following way. Let introduce the random variable $R_M^2=\C_M^2 +\S_M^2 $ (Fisher 1993).
From Eq.(\ref{def2a}) the upper limit now is $M$ and $\Phi_{\l m}=0$, one can have
\begin{equation}
R_M^2=\sum_{m,m^{'}}^M \cos(\Psi_{\l,m}-\Psi_{\l,m^{'}})= M +\sum_{m,m^{'}\neq m}^M \cos(\Psi_{\l,m}-\Psi_{\l,m^{'}}).
\label{R}
\end{equation}
For $M\gg 1$ and non-correlated phases the last term in Eq.(\ref{R}) is in order of $M^{-1}$ and $R_M^2=M$.

If the vector $G_M(\l)$ defines the distance from zero point, at what direction do the walkers move in the plane after
$M$ steps? The answer to this question can be obtained from circular statistics, namely, by $\Theta(\l,\Delta_\l)$.
Moreover, one can formulate the problem of two walks, which randomly move on the plane from the same initial
point $C$ and after $M$ steps come to some points $A$ and $B$. What is the probability to find both walkers at the vicinity of that points, if the distance between $A$ and $B$ is $r$?  More importantly, if the steps of two walkers are correlated, what is the distance between them after $M$ steps? The list of problems can be significantly extended, but using Pearson's walk model we can reformulate some quantities of the trigonometric statistics in terms of random walk approach. However, some of the statistical estimators can be easily derived using trigonometric moments, which we will demonstrate in the next section.
At the end of this Section we would like to point out that the properties of the $\Theta(m,\Delta_m)$ statistics can be described in term of random walk model, but now the "walker" made the steps
in $\l$ direction starting from $m=\l_{\max}$ and down to $m=1$ having corresponding variance
$R_m^2=\l_{\max}-m+1$. That sort of statistics allows us to investigate the morphology of the space of phases for each fixed value of $m$ in direction $m\le \l\le \l_{\max}$.

\section{Trigonometric statistics in application to WMAP data}
In this section we will investigate the asymmetry of different CMB maps derived from the WMAP data sets by Bennet et al. (2003) (ILC map), the Foreground Cleaned Map (FCM) and Wiener Filtered Map (WFM) by Tegmark et al. (2003), the Internal Linear Combination map reproduced by Erikson et al. (2004) (EILC). All these maps are derived by different methods of CMB and
foreground signal separation, applying different parts of the sky with masks and disjoint regions and they have different angular scales for the CMB image reconstruction. In the analysis which
follows, we are not going to discuss the limitation of the methods applied by the authors of ILC, FCM, WFM and EILC, or use these maps as possible illustration of different statistical properties of the signal. Our goal is
to show the sensitivity the trigonometric and Pearson's random walk statistics in application to different morphologies of the maps. For simplicity we will continue to use the term CMB map, but the reader should bear in mind that we use these maps for their illustrative character.

For example, the FCM derived with some smoothing by the Gaussian filter of the
signal at angular scale $\theta<1^\circ$, where the amplitudes (but not the phases!) of the signal is vanishing within that scale. The WFM
has some additional linear (Wiener) filter which was applied separately for each disjoint region of the FCM using the WMAP best-fit power spectrum $\cl$. The question is: how can we compare these maps, which have different
resolutions and different error bar level after foreground and noise separation?

\begin{figure}
\epsfig{file=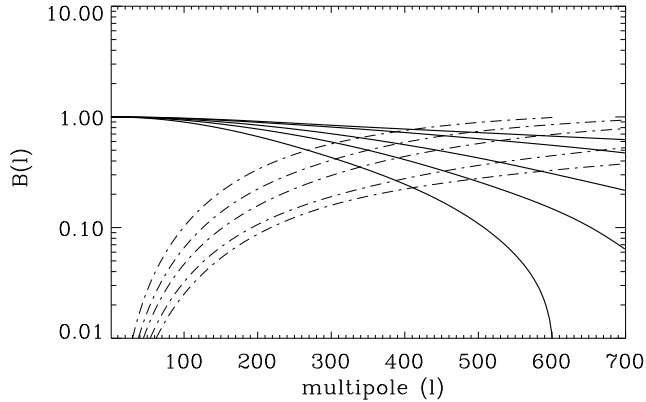,width=9cm}
\caption{The window function $w(\l)$ and $1-w(\l)$ functions for the K-W bands versus $\l$. The solid lines and dash-dotted lines are $w(\l)$ and $1-w(\l)$, respectively, from top to bottom for \wmap K, Ka, Q, V and W bands.}
\label{beam}
\end{figure}
One may argue that the best way to compare two maps is simply taking the
difference between 2 maps (pixel by pixel). However, due to different resolutions and different errors, taking the difference of 2 maps is by no means a good estimator. To elucidate this point, we show in Fig.\ref{beam} the function $w(\l)=W^{\frac{1}{2}}(\l)$, where $W(\l)$ is the window function for all \wmap K-W bands. The properties of the $w(\l)$ . In each band there are two parts of singal, $S_1$ and $S_2$, where $S_1$ is the sky signal $S_0$ convolved with the beam $B$: $S_1 = S_0 \bigotimes B$, and $S_2$ is the noise. The phases of the $\alm$ harmonic coefficients are therefore
\begin{equation}
\tan \Psi_{\l,m}^S=\frac{|{S_1}_{\l,m}|\sin\Psi^{S_1}_{\l,m} + |{S_2}_{\l,m}|\sin\Psi^{S_2}_{\l,m}}{
|{S_1}_{\l,m}|\cos\Psi^{S_1}_{\l,m} + |{S_2}_{\l,m}|\cos\Psi^{S_2}_{\l,m}}
\label{beam}
\end{equation}
If $|{S_1}_{\l,m}|\gg |{S_2}_{\l,m}|$, which is a good approximation for the low $\l$ range, the properties of the phase $\Psi_{\l,m}^S$ are the dominated by signal $S_1$. However,
when $|{S_1}_{\l,m}| \ll |{S_2}_{\l,m}|$ (for the high $\l$ range), the phases of the
signal $\Psi_{\l,m}^S$ are determined by the phases of the noise. Using these asymptotics,
we can conclude that the range of multipoles for which $|{S_1}_{\l,m}|\simeq |{S_2}_{\l,m}|$ is the range of transition ($\l \sim \l_{*}$) from the phases of the signal from the sky to the phases of the noise (see Chiang et al. 2002 for details). With different beams and window functions, the transition range is different for different bands.

\begin{figure}
\epsfig{file=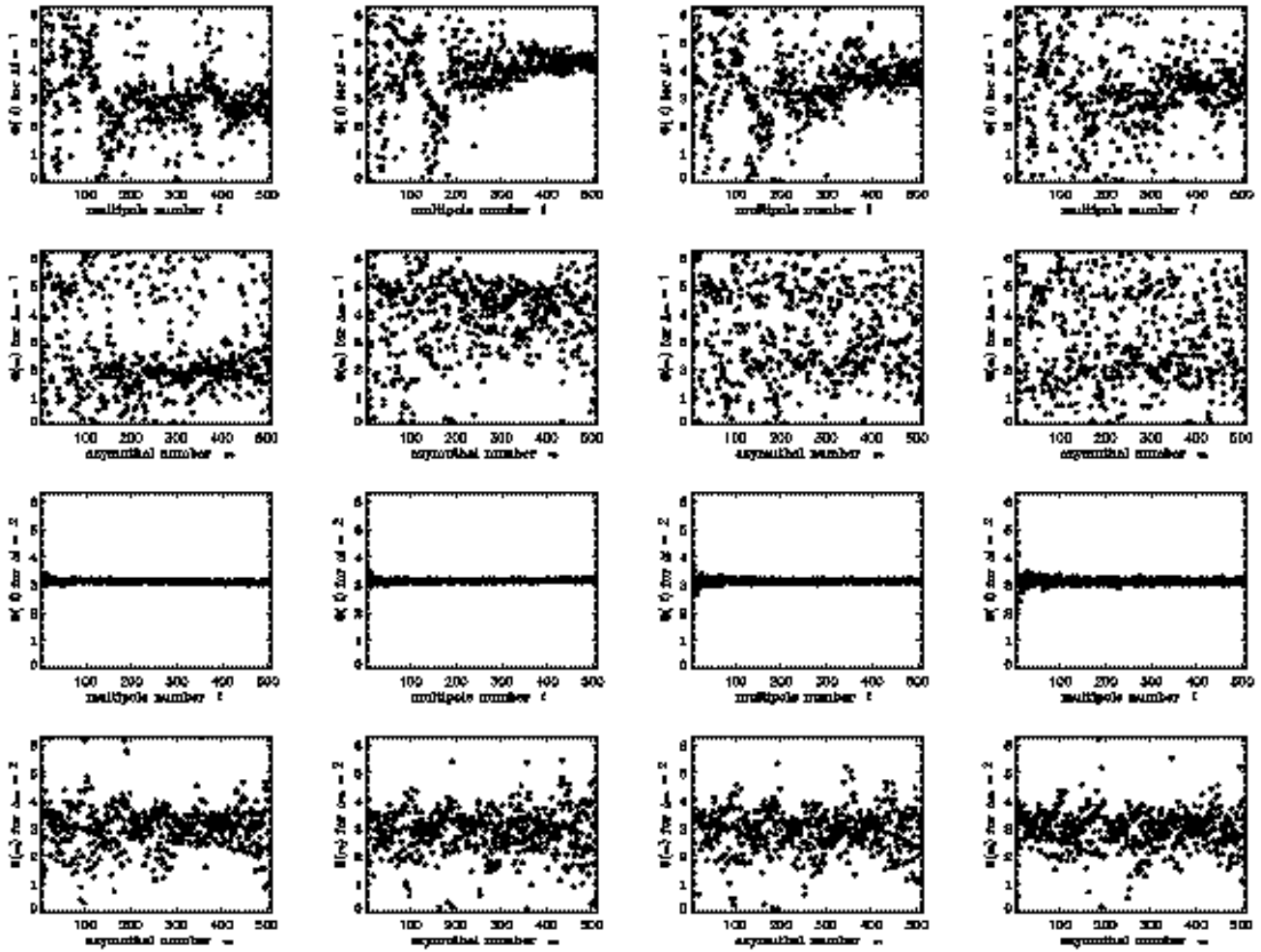,width=18cm}
\caption{Auto-correlation of phases in terms of $\Theta_a(\l,\Delta_\l)$, $\Theta_a(m,\Delta_m)$ for the \wmap (from left to right columns) Ka, Q, V and W bands. From top to bottom are $\Delta_\l=1$ and $\Delta_m=1$, $\Delta_\l=2$ and $\Delta_m=2$, respectively.}
\label{wmapauto}
\end{figure}

We propose another way for comparison of the different CMB and non-CMB maps, based on asymmetry parameters. Namely, for each map, using coefficients of the spherical harmonics expansion, we derive the phases $\phi^j_{\l,m}$ for comparison using auto- and cross-correlation of the phases. Below we present some of the results of the analysis. We include in our analysis not only phases of the derived CMB maps, but phases of the K-W bands of the \wmap data sets in order to compare the properties of the CMB and
original data. In Fig.\ref{wmapauto} we plot the trigonometric moments for the phases of the \wmap Ka-W band signals. These are supposedly pronounced non-Gaussian signals due to the Galactic contaminations, and $\Theta_a(\l,\Delta_\l)$, $\Theta_a(m,\Delta_m)$ display the highly concentrated regions, in particular for $\delta_\l =2$, therefore indeed non-Gaussian.

\section {Phase auto-correlations for the derived CMB maps}
In this section we discuss the properties of the CMB maps derived from the \wmap data sets by combination of the K-W bands maps. As is mentioned in Introduction we use the ILC, FCM, WFM and EILC and the corresponding $\alm$
coefficients for all range of multipoles ($\l_{\max}=512$). However, none of these
maps represent pure CMB signal up to this range of $\l$ because of different restrictions of the methods applied for separation of the CMB signal, noise and foregrounds. Meanwhile, these maps allows to test the properties of the trigonometric statistics of the phases, which as we believe can be useful for testing the upcoming \wmap data and the \planck data.

The statistics we use are $\Theta_a(\l,\Delta_\l)$, $ \Theta(m,\Delta_m)$ for
$\Delta_\l=1,2$ and $\Delta_m=1,2$.

\begin{figure}
\epsfig{file=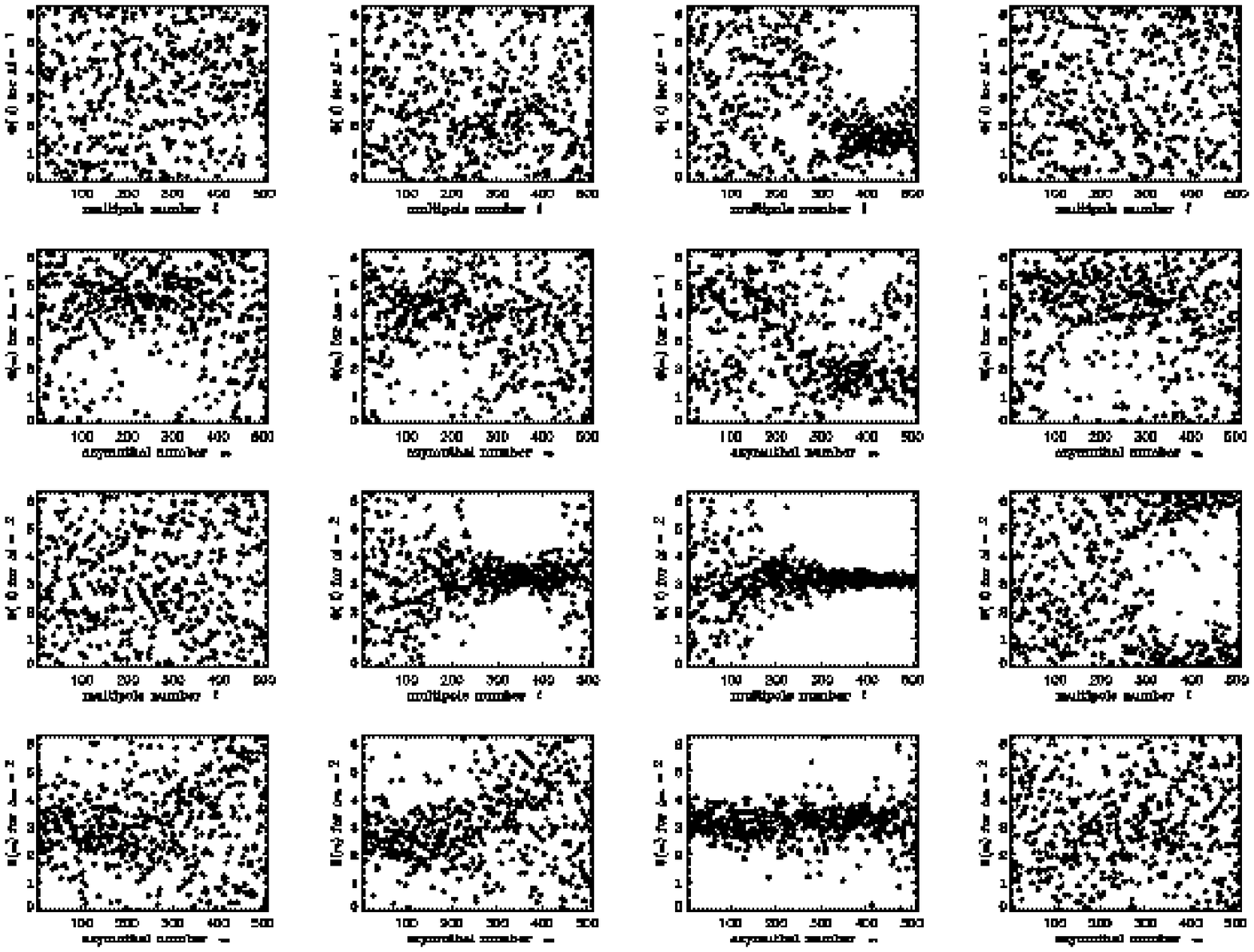,width=18cm}
\caption{Auto-correlation of phases in terms of $\Theta_a(\l,\Delta_\l)$, $ \Theta_a(m,\Delta_m)$ functions from left to right columns for the ILC, EILC, FCM and WFM, respectively. From top to bottom rows are $\Delta_\l=1$,  $\Delta_m=1$, $\Delta_\l=2$ and $\Delta_m=2$, respectively.}
\label{cmbauto}
\end{figure}

For the random signal all these functions are uniformly distributed by definition at the range $[0,2\pi]$, while for the non-Gaussian signal they should display clusterization of the phases. To compare between the phases of maps, in Fig.\ref{cmbauto} we plot $\Theta(\l,\Delta_\l)$, $\Theta(m,\Delta_m)$ functions from left to right ILC, EILC, FCM, WFM, respectively.
We also would like to draw attention on the morphology of the $\Theta_a(\l,\Delta_\l)$, $ \Theta(m,\Delta_m)$ functions for $\Delta_\l=1,2$ and $\Delta_m=1,2$. The correlations for the WFM are discussed in Naselsky et al. 2004 and are independently found by Prunet et al. (2004) for the \wmap signal. However, for the ILC map the $\Theta(m,\Delta_m)$ statistics have displayed some non-Gaussian character, where for $\Delta_m =1$, $\Theta$ are more clustered between $\pi$ and $2\pi$ and for $\Delta_m =2$, it is more concentrated around $\pi$. 

\section{Cross-correlations between the derived CMB maps}
\begin{figure}
\epsfig{file=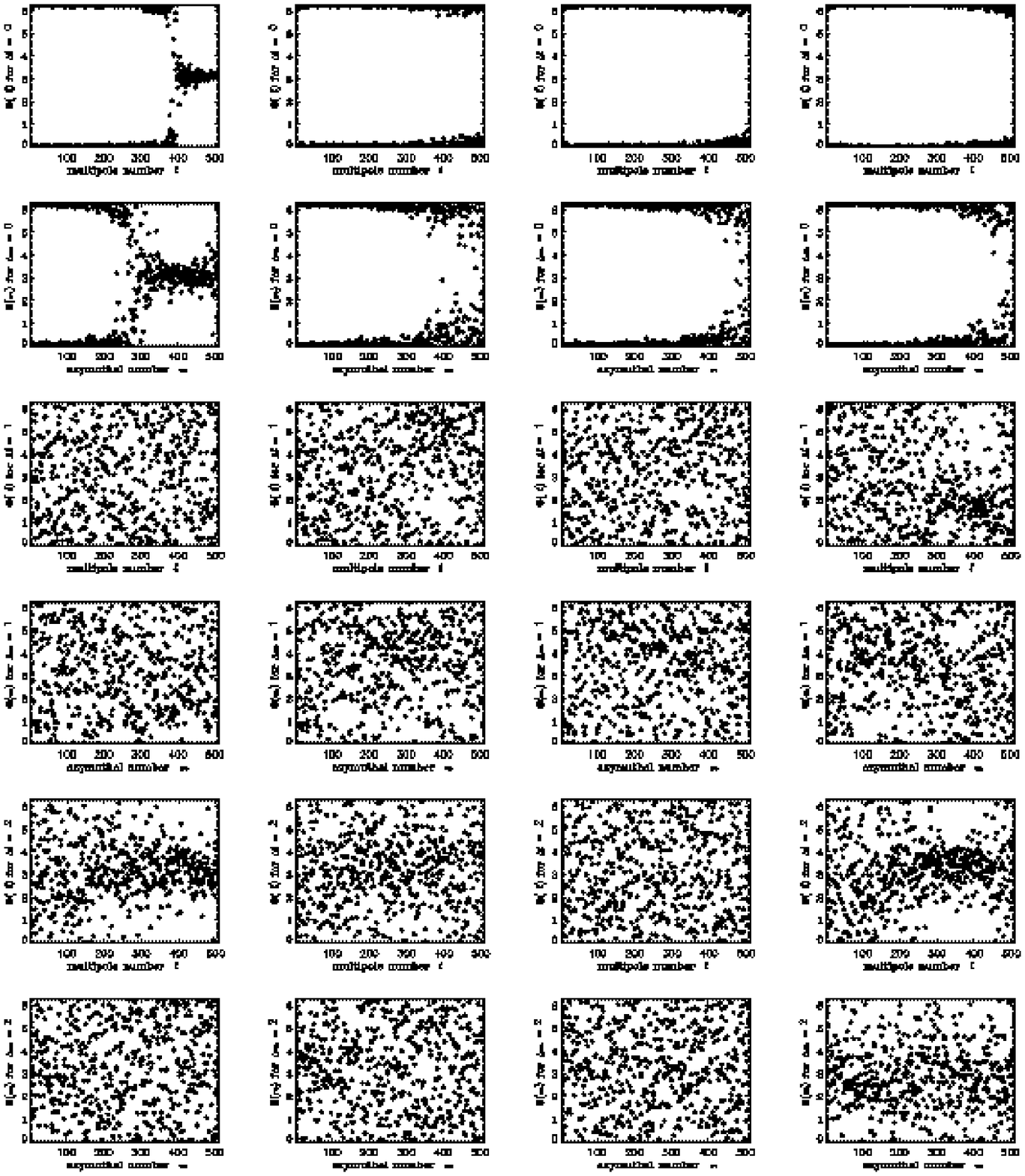,width=18cm}
\caption{Cross correlation of phases in terms of $\Theta_c(\l,\Delta_\l)$, $ \Theta_c(m,\Delta_m)$ functions between a pair of maps. From left to right columns are between ILC and EILC, ILC and FCM, ILC and WFM, and between FCM and EILC. From top to bottom rows are  $\Delta_\l=0$, $\Delta_m=0$, $\Delta_\l=1$, $\Delta_m=1$, $\Delta_\l=2$ and $\Delta_m=2$,  respectively.}
\label{cmbcross}
\end{figure}

In this Section we would like to demonstrate how $\Theta_c(\l,\Delta_\l)$, $\Theta_c(m,\Delta_m)$ statistics can be
used for comparison of different CMB maps derived by different methods, including the linear (Gaussian or Wiener) filters. These linear filters significantly destroy some information at specific ranges of multipoles, but not the phases of the signals. Convolving with a Gaussian filter with characteristic scale $\l=\l_G$, all the amplitudes $\alm$ for $\l\gg \l_G$ become vanishing  while all the phases preserve all information
about the morphology of the CMB map. That is one of the reasons why phase analysis is essential for testing of the statistical properties of the derived CMB maps, as it clearly illustrates different outcomes of the cleaning methods used. To compare the phases of ILC, EILC, FCM and WFM we implement Eq.(\ref{def2a}), in which the phase $\Psi$ now corresponds to one map (for example, ILC) and the phase $\Phi$ corresponds to another (for example, FCM). We exploit the simplest fact that if the phases are the same, in terms of $\Theta_c(\l,\Delta_\l)$, $\Theta_c(m,\Delta_m)$ statistics, they behave like strong non-Gaussian signal. In Fig.\ref{cmbcross} we plot the cross-correlators $\Theta_c(\l,\Delta_\l)$, $ \Theta_c(m,\Delta_m)$ by pairing CMB maps: from left to right are between ILC and EILC, ILC and FCM, ILC and WFM, and between EILC and FCM. 


Comparing the phases of ILC and EILC maps one can see significant $\pm \pi$ anticorrelations for $\l \geq 400$ at $\Delta \l=0$, while for other pairs they are negligible. And  at $\Delta m=0$ all pairs start deviating from total correlation for $m \geq 300$, indicating that the main difference in morphology of these derived maps starts from the multipole range $\l>400$. More interestingly, for the ILC and EILC maps the phase correlations have systematic shift by $\pm \pi$. Note that while in terms of $\Theta_c(\l,\Delta_\l=0)$ and $ \Theta_c(m,\Delta_m=0)$ statistics the signals are nearly the same for $\l \leq 400$, they have different auto-correlation (see Fig.\ref{cmbauto}), which is why the $\Theta_c(\l,\Delta_\l=1,2)$ and $\Theta_c(m,\Delta_m=1,2)$ statistics are different. 

\begin{figure}
\epsfig{file=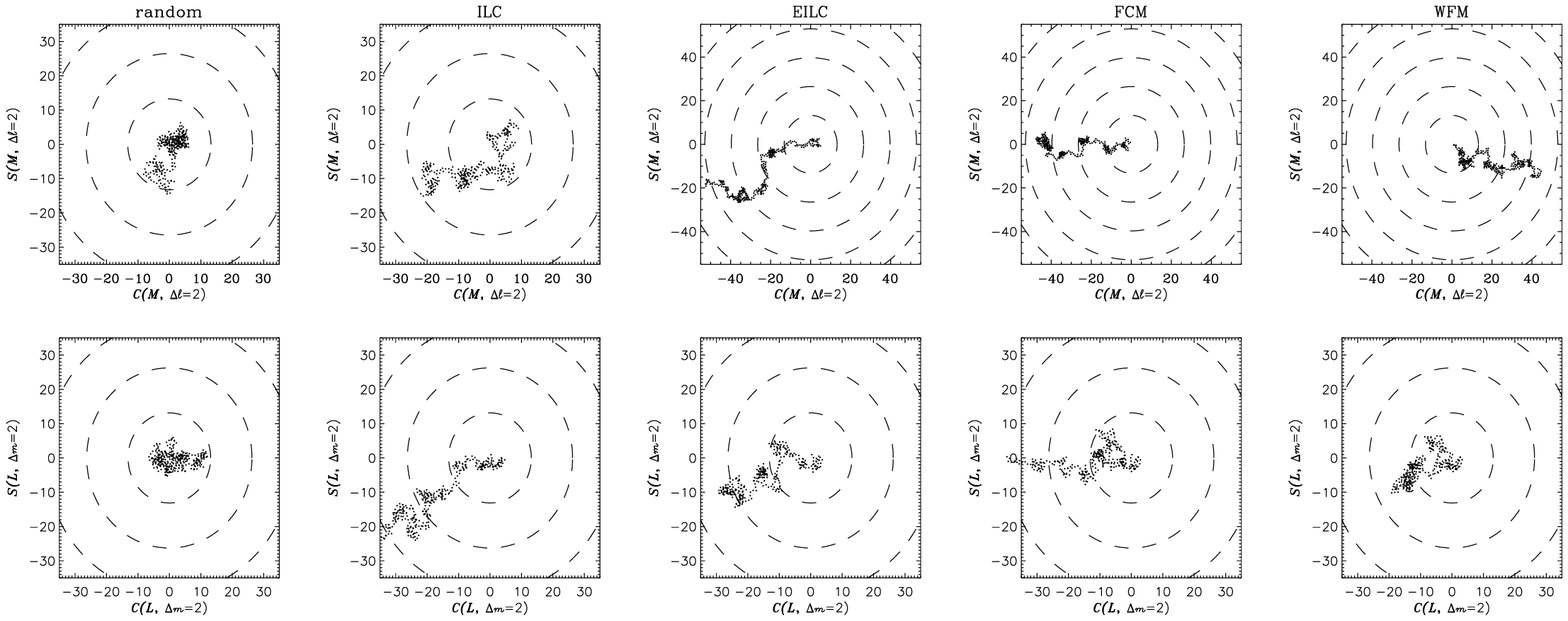,width=18cm,height=6.5cm}
\caption{Random walk of $\sum^M \exp(i \Delta \Phi)$. In each panel the $x$ and $y$ axes are $\C(M)$ and $\S(M)$ (top row), and $\C(L)$ and $\S(L)$ (bottom), respectively and the circles (starting from inner one) represent $1, 2, 3, 4 \sigma \ldots$ of Gaussian confidence levels for maximum $M_{\max}=350$ and $L_{\max}=350$. The top row is for $\Delta_\l=2$ of $\l=350$ and bottom for $\Delta_m=2$ of $m=6$ with $L_{\max}=350$. From left to right columns are Gaussian random signal, ILC, EILC, FCM and WFM, respectively.}
\label{rwalk}
\end{figure}

\section{Deviation from randomness in terms of Pearson's walk statistics}
As is mentioned in Section III, the properties of the trigonometric moments can be understood in terms of the Pearson's random walk. It provides some additional information about auto and cross-correlation of phases even if the deviation after $M$ steps from the estimated 
$R_M$ is $2-3$ times bigger than for the Gaussian estimator $R_M=M^{1/2}$. Although the $\Theta_c(\l,\Delta_\l)$ and $\Theta_c(m,\Delta_m)$ functions show the gross behavior for a certain separation, the random walk of phases can display how the phases correlate step by step. 

In Fig.\ref{rwalk}, we present the random walk for Gaussian random signal, ILC, FCM and EILC with fixed separation $\Delta \l=2$ for $\l=350$ and $\Delta m=2$ for $m=6$. We choose the particular case of $\Delta \l=2$ for $\l=350$ because the resultant angles for EILC and FCM are close to $\pi$, as shown in Fig.\ref{cmbauto} the 3rd row, the same reason for $\Delta m=2$ for $m=6$ for ILC and FCM. The Gaussian $\sigma$ values are calculated when we set the $L_{\max}=350$ and $M_{\max}=350$. The walks projected in two axes can be written as
\begin{eqnarray}
\C_{\l=350,\Delta_\l=2}(M)= \sum_{m=1}^M \cos(\Psi_{\l+\Delta_\l,m}-\Psi_{\l,m})\nonumber\\
\S_{\l=350,\Delta_\l=2}(M)= \sum_{m=1}^M \sin(\Psi_{\l+\Delta_\l,m}-\Psi_{\l,m})
\label{pear}
\end{eqnarray}
and
\begin{eqnarray}
\C_{m=6,\Delta_m=2}(L)=\sum_{\l=m}^{L} \cos(\Psi_{\l,m}-\Psi_{\l,m-\Delta_m})\nonumber\\
\S_{m=6,\Delta_m=2}(L)= \sum_{\l=m}^{L} \sin(\Psi_{\l,m}-\Psi_{\l,m-\Delta_m}).
\label{pearrr}
\end{eqnarray}

As one can see from Fig.\ref{rwalk} the walks of the derived CMB maps are intrinsically different from the random signal. For ILC, the walk with the phase difference $\Delta m=2$ reaches $3\sigma$, so do the walks with $\Delta \l=2$ for EILC, FCM and WFM. Although the $\Theta$ statistics from trigonometric moments cannot provide any information about non-Gaussianity for a single $\l$ or $m$, through Pearson's random walk we can see significant correlation of the phases in any chosen direction. The corresponding values of the $R_M$ parameter for the maps are around or above $3\sigma$.

\section{Rotational invariance of the trigonometric moments of phases}
The previous analysis of the trigonometric moments as estimators of non-Gaussianity of the
anisotropy and polarization maps dealing with phases of $\alm$ coefficients consequently depends on the reference system of coordinates. Obviously, for different coordinate systems these $\alm$ coefficients and the corresponding phases are different. The issue which we would like to discuss in this section is how phase correlations depend on the reference system of coordinates (M. Hobson, private communication, see also Coles et al. 2003). We concentrate on the following question: are the phases correlators, in particular, the trigonometric moments we introduce, rotationally invariant? And if not, how significantly such non-invariance can transform the conclusion about non-Gaussianity of the maps? To answer these questions we need to know how to transform a given set of $\alm$ of a given coordinate system to new coefficients $b_{\lm}$,  which corresponds to the new coordinate system rotated by the Euler angles $\alpha, \beta, \gamma$. Following the general method (Varshalovich, Moskalev and Khersonskii 1988) and taking into account properties of the spherical harmonics (see Coles et al. 2003) we get
\begin{equation} 
 b_{\l,m}=\sum_{m^{'}} D^\l_{m,m^{'}}(\alpha,\beta,\gamma)a_{\l,m^{'}}
\label{rot1}
\end{equation}
where $D^\l_{m,m^{'}}(\alpha,\beta,\gamma)$ is the spherical harmonic decomposition 
of the Wigner function $\textbf{D}(\alpha,\beta,\gamma)$. The coefficients 
$D^\l_{m,m^{'}}(\alpha,\beta,\gamma)$ should preserve the moduli $\sum_m |b_{\l,m}|^2=
\sum_m|a_{\l,m}|^2$ under transformation. This leads to the following equations for 
$D^\l_{m,m^{'}}(\alpha,\beta,\gamma)$
\begin{equation} 
 \sum_m|b_{\l,m}|^2=\sum_m\sum_{m^{'},m^{"}} D^\l_{m,m^{'}}(\alpha,\beta,\gamma)
 {D^{*}}^\l_{m,m^{"}}(\alpha,\beta,\gamma)a_{\l,m^{'}}{a^{*}}_{\l,m^{"}}=\sum_m|a_{\l,m}|^2
\label{rot2}
\end{equation}
and
\begin{equation}
\sum_m D^\l_{m,m^{'}}(\alpha,\beta,\gamma)
 {D^{*}}^\l_{m,m^{"}}(\alpha,\beta,\gamma)=\delta_{m^{'},m^{"}}
\label{rot3}
\end{equation}
where $\delta_{m^{'},m^{"}}$ is the Kroneker $\delta$ symbol.
Without loss of generality we assume that for the initial reference system of coordinate 
$a_{\l,m}=\exp(i\Phi_{\l,m})$ where $\Phi_{\l,m}$ is the phase for a given $\l$ and $|m|\le \l$. To obtain the trigonometric moments for the system after rotation we define a matrix of correlations: 
\begin{equation}
G_{\l,\l^{'}}=\sum_m b_{\l,m}b^{*}_{\l^{'},m}=\sum_m\sum_{m^{'},m^{"}} 
D^\l_{m,m^{'}}(\alpha,\beta,\gamma)
D^{*\l^{'}}_{m,m^{"}}(\alpha,\beta,\gamma)a_{\l,m^{'}}a^{*}_{\l^{'},m^{"}}
\end{equation}
where $\l^{'}=\l+\Delta_\l$ and $b_{\lm}=|b_{\l,m}|\exp(i\Psi_{\l,m})$, and $\Psi_{\l,m})$ is the phase of the $\l,m$ harmonics after rotation. For the statistical ensemble of realizations of the same Gaussian random process, average over realizations leads to orthogonality of the $\a_{\lm}$ coefficients:
$\langle a_{\l,m^{'}}a^{*}_{l^{'},m^{"}} \rangle = \delta_{m^{'},m^{"}}\delta_{\l,\l^{'}}$. 
Thus, the correlation matrix has the form $G_{\l,\l^{'}}=\delta_{\l,\l^{'}}$ which is
typical for a Gaussian process. Unfortunately, in CMB studies we are dealing with one single realization of the sky, therefore such asymptotics can not be achieved. In a statistical sense, however, non-correlated phases in a Gaussian signal preserve randomness of the trigonometric moments and could play a role for estimation of any non-Gaussianity of the map.

Let's discuss the properties of trigonometric moments for rotated maps. From Eq.(\ref{rot1}) we obtain
\begin{eqnarray}
\Theta(\l,\alpha,\beta,\gamma)= \tan^{-1}\frac{\Im \left\{ \sum_m^{} \sum_{m^{'},m^{"}} D^\l_{m,m^{'}}(\alpha,\beta,\gamma) D^{*\l^{'}}_{m,m^{"}}(\alpha,\beta,\gamma) \exp[i(\Phi_{\l,m^{'}}-\Phi_{\l^{'},m^{"}})]\right\}}
                                              {\Re \left\{ \sum_m \sum_{m^{'},m^{"}} D^\l_{m,m^{'}}(\alpha,\beta,\gamma) D^{*\l^{'}}_{m,m^{"}}(\alpha,\beta,\gamma) \exp[i(\Phi_{\l,m^{'}}-\Phi_{\l^{'},m^{"}})]\right\}}
\label{rot5}
\end{eqnarray}

\begin{figure}
\epsfig{file=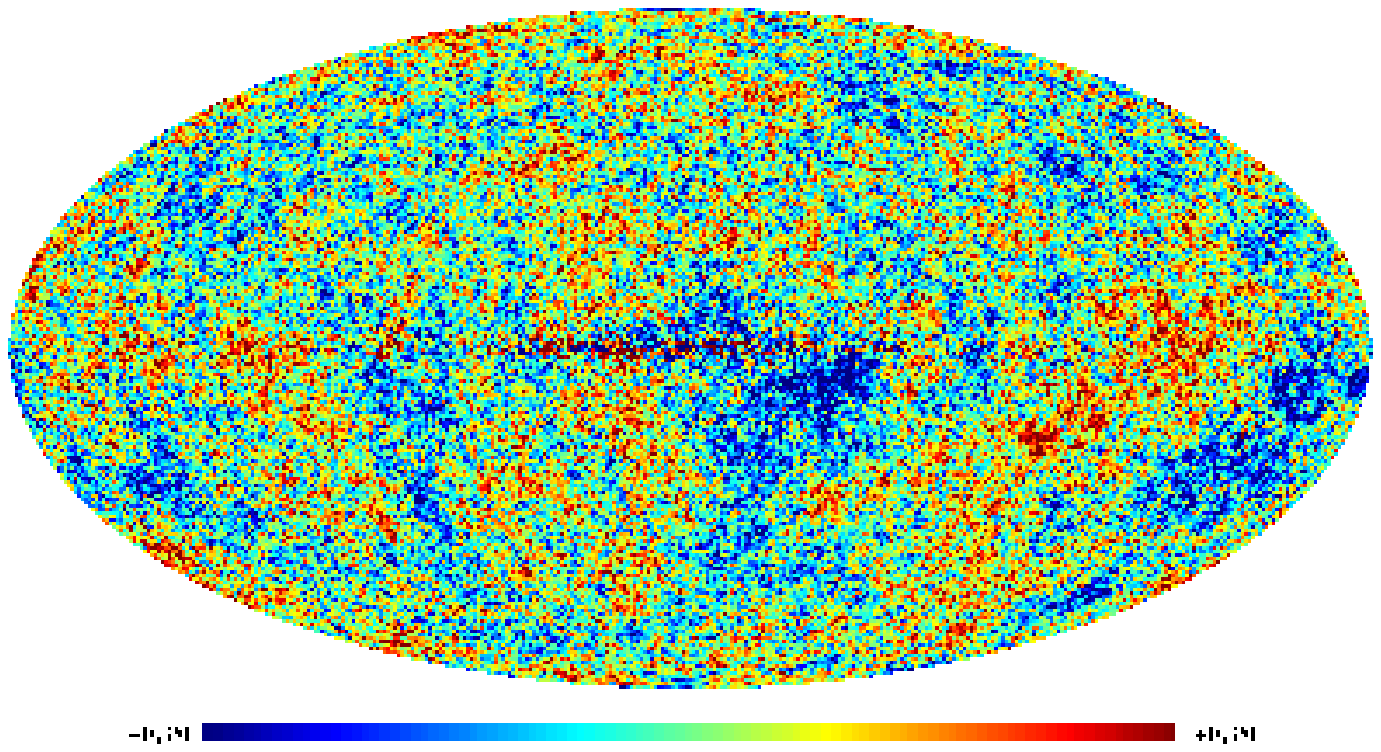,width=8cm}
\epsfig{file=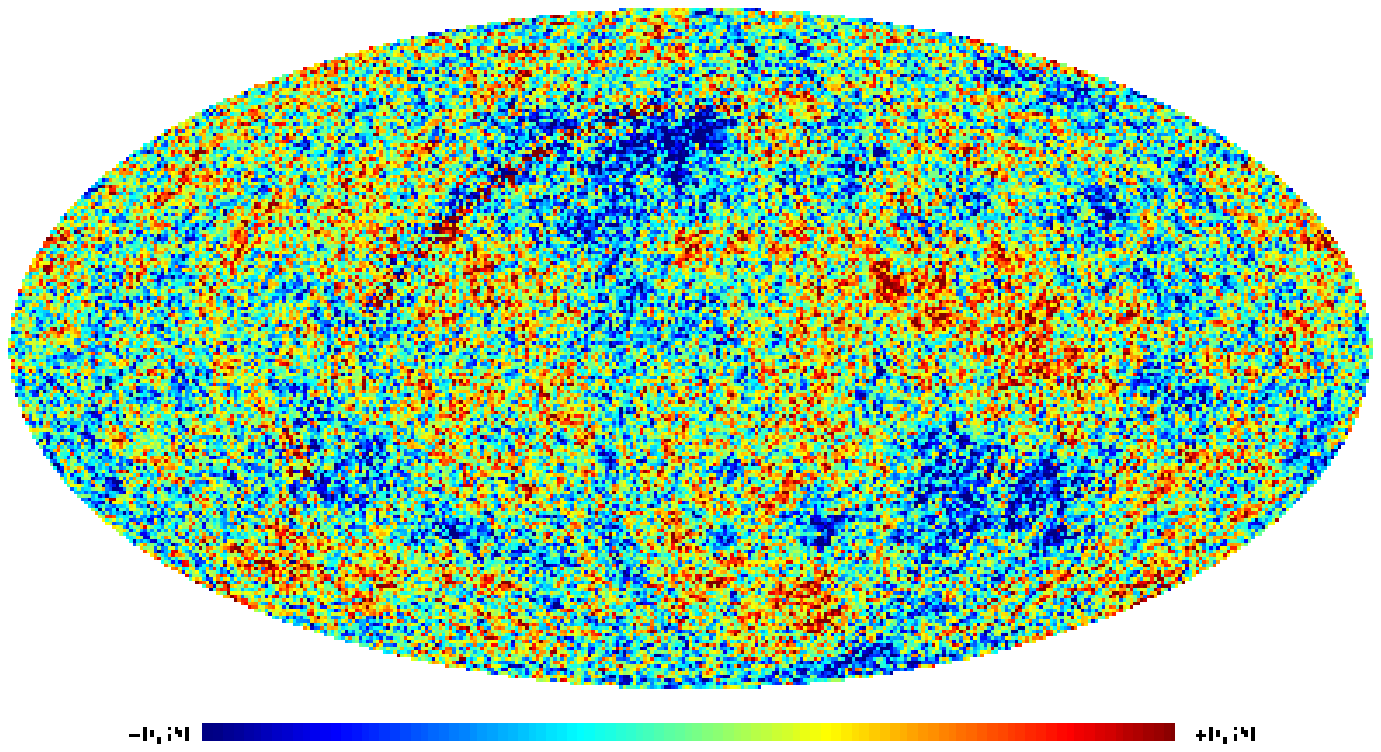,width=8cm}
\caption{Rotation of the maps. The left is the FCM map and the right is its rotated map by $\theta=55^\circ$ and $\phi=40^\circ$.}
\label{rot}
\end{figure}

Taking into account (see Varshalovich, Moskalenko and Khersonskii, 1988)
\begin{equation}
D^\l_{m,m^{'}}=\exp(-im\alpha)d^\l_{m,m^{'}}(\beta)\exp(-i m^{'}\gamma);\nonumber\\
\label{rot6}
\end{equation}
where
\begin{eqnarray}
d^\l_{m,m^{'}}(\beta)=\left[(\l+m^{'})!(\l-m^{'})!(\l+m)!(\l-m)!\right]^{\frac{1}{2}}
\sum_k^{\min(\l-m,\l-m^{'})} (-1)^{\l^{'}-m^{'}-k}\frac{(\l-m-k)!k!}{(\l-m^{'}-k)!(k+m+m^{'})!}\nonumber\\
\times(\cos\beta)^{2k+m+m^{'}}(\sin\beta)^{2\l-2k-m-m^{'}}
\label{rot7}
\end{eqnarray}
we obtain the following formula for $ \Theta(\l,\alpha,\beta,\gamma)$:
\begin{eqnarray}
\Theta(\l,\alpha,\beta,\gamma)=\tan^{-1}\frac{\Im\left\{ \sum_{m^{'},m^{"}} W^{\l,\l^{'}}_{m^{'},m^{"}}(\beta)
\sin\left(\Phi_{\l,m^{'}}-\Phi_{\l^{'},m^{"}}+\gamma(m^{'}-m^{"})\right)\right\}}
{\Re\left\{\sum_{m^{'},m^{"}} W^{\l,\l^{'}}_{m^{'},m^{"}}(\beta)
\cos\left(\Phi_{\l,m^{'}} - \Phi_{\l^{'},m^{"}} + \gamma(m^{'}-m^{"})\right)\right\}}
\label{rot8}
\end{eqnarray}
where
\begin{eqnarray}
W^{\l,\l^{'}}_{m^{'},m^{"}}(\beta)=\sum_m d^\l_{m,m^{'}}(\beta)d^{\l^{'}}_{m,m^{"}}(\beta)
\end{eqnarray}
As one can see from Eq.(\ref{rot8}) the $\Theta(\l,\alpha,\beta,\gamma)$ statistic is
rotationally invariant if $\beta=0$ and $\gamma=0$. It is practically rotationally
invariant even when $\beta=0$ but $\gamma \neq 0$. For that case as Eq.(\ref{rot8}) the main contribution to the summation in both the nominator and denominator corresponds to $m^{'}=m^{"}$ and $\Theta(\l,\alpha,\beta=0,\gamma) \simeq \Theta(\l,\alpha=0,\beta=0,\gamma=0) $, where
$\Theta(\l,\alpha=0,\beta=0,\gamma=0)$ corresponds to the initial reference system.
However, if $\beta\neq 0$, the rotational invariance is broken and the statistical properties
of $\Theta(\l,\alpha,\beta,\gamma)$ and $\Theta(\l,\alpha=0,\beta=0,\gamma=0)$ are different
for given multipoles $\l,m$. To investigate how significantly the $ \Theta(\l,\alpha,\beta,\gamma)$ 
statistics can be changed we resort to the Monte Carlo simulation presented below.
We use the GLESP code (Doroshkevich et al. 2003), which provides map rotation for arbitrary angles $\theta$ and $\phi$. In Fig.\ref{rot} we show the rotation
of the FCM map by $\theta=55^\circ$ and $\phi=40^\circ$, and calculate the trigonometric moments shown in Fig.\ref{rot1}. 

\begin{figure}
\epsfig{file=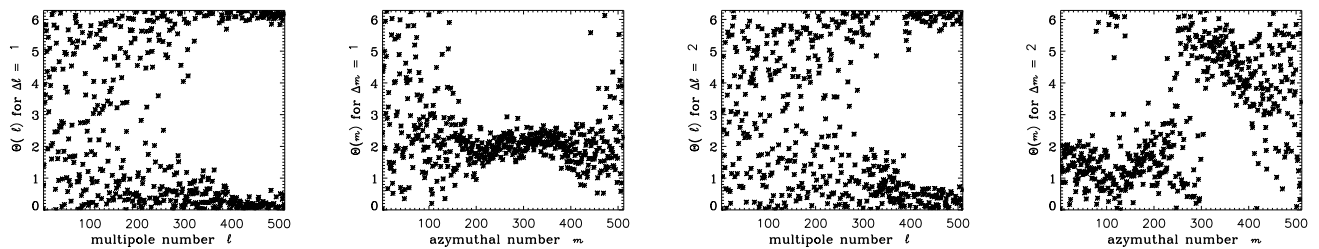,width=18cm}
\caption{The mean angle statistics for auto correlations for the rotated FCM map as shown in Fig.\ref{rot}. From left to right are $\Delta_\l=1$, $\Delta_m=1$, $\Delta_\l=2$ and $\Delta_m=2$, respectively. These are to be compared with the 3rd column in Fig.\ref{cmbauto}.}
\label{rot1} 
\end{figure}
As one can see from Fig.\ref{rot1}, the non-Gaussianity of the FCM map is preserved in different form in terms of the mean angle statistics $\Theta(\l,\alpha,\beta,\gamma)$. The residuals from the Galactic plane contributes
significantly to $ \Theta(\l,\alpha=0,\beta=0,\gamma=0)$ statistics, starting from high multipoles ($\l \geq 200$). After rotation, the Galactic plane is shifted causing more correlations for $\Delta m=1$. In fact, non-invariance of phases provides a new way to test any residues, including from the Galactic plane, by rotation of the map by $\theta=\pi/2$, for
which any alignment in the structure in the original orientation is now perpendicular to the previous one. Once again we would like to emphasize that by definition for pure Gaussian signal any rotations should not lead to correlation of phases, which can be used for estimation of significance level of the non-Gaussianity.

\section{Conclusions}
In this paper we have presented a new method for analysis of non-Gaussianity for CMB maps.  is further development from the trigonometric moments of phases of the spherical harmonic coefficients $\alm$ can be concluded to form mean angles that enables us to see the global behavior of the ``randomness'' of the phases. Furthermore, the trigonometric moments are closely related to the Pearson's random walks and the mean angle is the resultant angle of the 2 dimensional random walk on the complex plane. The properties of such random walks are well-developed and they are used to display the phase correlations step by step.
We apply these analyses on the derived CMB maps from 1-year \wmap data. These CMB maps are derived from different foreground cleaning methods, hence their morphologies are somewhat different. As phases are closely related to morphology, our analyses on phases not only  demonstrate the existence of non-Gaussian residuals among these CMB maps, but also reveal the different morphologies of these maps.
With the upcoming ESA \planck mission, CMB maps with higher resolution and sensitivity will be derived by different foreground cleaning methods, for which testing the Gaussianity will be imminent. The methods and analyses we present in this paper for the \wmap data can be applied straightaway to new datasets by the \planck mission.

\section{Acknowledgments}
We thank Mike Hobson for useful discussions. We
acknowledge  the use of the Legacy Archive for Microwave Background
Data Analysis (LAMBDA) and the maps provided by Max Tegmark et al. We also
acknowledge the use of H{\sc ealpix} package \cite{healpix} and GLESP package \cite{glesp}.

%
%
\newcommand{\autetal}[2]{{#2.\ #1 {\it et al}.,}}
\newcommand{\aut}[2]{{#2.\ #1,}}
\newcommand{\saut}[2]{{#2.\ #1,}}
\newcommand{\laut}[2]{{and #2.\ #1,}}
\newcommand{\psaut}[2]{{#2.\ #1}}

%
%
\newcommand{\refs}[6]{#2 {\bf #3}, #4 (#5).}
\newcommand{\midrefs}[6]{#2 {\bf #3}, #4 (#5);}
\newcommand{\unrefs}[6]{#2 #3.}
\newcommand{\midunrefs}[6]{#2, #3;}

%
\newcommand{\book}[6]{{\it #1} (#2\, #5).} 
\newcommand{\midbook}[6]{{\it #1} (#2\, #5);} 

%
\newcommand{\proceeding}[6]{#5, in #4, edited by #3 (#2#5).} 
\newcommand{\midproceeding}[6]{#5, in #4, edited by #3 (#2#5);} 

\newcommand{\combib}[3]{\bibitem{#3}} 

%
%
\def\apjl{Astrophys.\ J.\ Lett.}
\def\mn{Mon.\ Not.\ R.\ Astron.\ Soc.}
\def\nat{Nature (London)}
\def\araa{Ann. Rev. Astron. Astrophys.}
\def\aa{Astron.\ Astrophys.}
\def\apj{Astrophys.\ J.}
\def\apjs{Astrophys.\ J. Supp.}
\def\prl{Phys.\ Rev.\ Lett.}
\def\prd{Phys.\ Rev.\ D}
\def\pl{Phys.\ Lett.}
\def\np{Nucl.\ Phys.}
\def\rmp{Rev.\ Mod.\ Phys.}
\def\cmp{Comm.\ Math.\ Phys.}
\def\mpl{Mod.\ Phys.\ Lett.}
\def\pr{Phys. Rep.}
\def\ijmpd{Int. J. Mod. Phys. D}

\newcommand{\amp}{\& }
\def\cqg{Class.\ Quant.\ Grav.}
\def\grg{Gen.\ Rel.\ Grav.}

\def\cup{Cambridge University Press, UK}
\def\princetonpress{Princeton University Press}
\def\worldpress{World Scientific, Singapore}
\def\oxfordpress{Oxford University Press}

\end{document}